\documentclass[apl,twocolumn,showpacs,amsmath,letter]{revtex4}
\usepackage{graphicx}

\newcommand{\Journal}[4]{#1 \textbf{#2}, #3 (#4)}

\begin{document}

\title{Thermal Dynamics in Symmetric Magnetic Nanopillars Driven by Spin Transfer}

\author{Weng L. Lim}
\author{Nicholas Anthony}
\author{Andrew Higgins}
\author{Sergei Urazhdin}
\affiliation{Department of Physics, West Virginia University, WV
26506}


\begin{abstract}

We study the effects of spin transfer on thermally activated
dynamics of magnetic nanopillars with identical thicknesses of the
magnetic layers. The symmetric nanopillars exhibit anomalous
dependencies of switching statistics on magnetic field and current.
We interpret our data in terms of simultaneous current-induced
excitation of both layers. We also find evidence for coupling
between the fluctuations of the layers due to the spin transfer.

\end{abstract}

\maketitle

Spin polarized current flowing through a nanoscale magnet F$_1$ can
change its magnetic configuration due to the spin transfer torque
(ST)~\cite{slonczewski,berger}. Real devices require an additional
nanomagnet F$_2$ to polarize the current flowing through F$_1$. To
minimize the complications due to the ST exerted on the polarizing
layer F$_2$, it is usually made much thicker than F$_1$, or pinned
by an adjacent antiferromagnet via exchange bias. In this case, the
magnitude of the current required to induce dynamics in F$_2$ is
expected to be much larger than in F$_1$.  However, simultaneous
effects of ST on several magnetic layers are important in many
devices such as those incorporating artificial antiferromagnets,
which are comprised of two or more magnetic layers of similar
thickness. Understanding these effects is important for enhancing
the performance and stability of magnetoelectronic devices.

In a symmetric bilayer comprised of two separate magnetic layers
F$_1$ and F$_2$ with similar dimensions, the direction of ST exerted
on both layers is the same. Consequently, ST has been suggested to
induce their coupled precession in a propeller-like fashion
~\cite{slonczewski}. Measurements of magnetoresistance (MR) in
symmetric Co-based nanopillars showed resistance increases at large
magnetic field $H$ for both directions of the applied current
$I$~\cite{tsoisymmetric,kentsymmetric}. These features were
attributed to current-induced precession of F$_1$ for one direction
of current, and of F$_2$ for the opposite direction. However, the
relation of these results to the theoretical picture of ST is not
clear.

In this Letter, we present measurements of the effect of ST on
symmetric nanopillars in thermally activated regime, i.e. when the
magnetic moments flip their directions randomly due to thermal
fluctuations enhanced by ST. We show that the dynamics of both
layers must be activated due to the simultaneous effect of ST on
both layers. We develop a simple activation model, and present data
indicating coupling between the dynamics of the two layers mediated
by ST.

Our samples with structure
F$_1$=Ni$_{80}$Fe$_{20}$=Py(4)/Cu(3)/F$_2$=Py(4) were deposited at
room temperature (RT) by magnetron sputtering at base pressure of
$5\times 10^{-9}$~Torr, in $5$~mTorr of purified Ar. All thicknesses
are in nm. The structure was patterned into a nanopillar with
dimensions $120\times 60$~nm sandwiched between Cu leads by a
process described elsewhere~\cite{myapl}. The patterning procedure
resulted in slightly smaller dimensions of $F_2$ than $F_1$, as
shown by measurements of MR. This slight asymmetry is important for
some of the behaviors described below. We measured $dV/dI$ at RT
with four-probes and lock-in detection, adding an ac current of
amplitude not exceeding 100~$\mu$A at 1~kHz to the dc current $I$.
Positive $I$ flows from $F_1$ to $F_2$. $H$ is in the film plane and
along the nanopillar easy axis. We report results for one of the
four samples tested with similar results.

\begin{figure}
\includegraphics[scale=0.8]{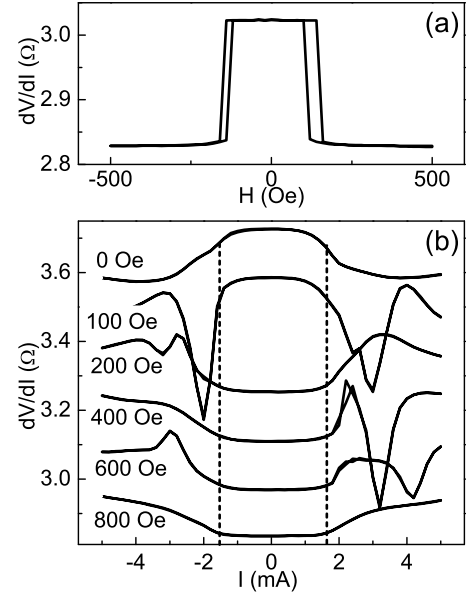}
\caption{\label{fig1} (a) $dV/dI$ {\it vs.} $H$ at $I=0$. (b)
$dV/dI$ {\it vs.} $I$ at $H$ as labeled. The curves are offset for
clarity.}
\end{figure}

In scans of field $H$ at $I=0$, the sample exhibits a nearly
reversible transition from antiparallel (AP) state with resistance
$R_{AP}$ at small $H$, to parallel (P) state with resistance $R_P$,
at field $H_d\approx 137$~Oe characterizing the dipolar coupling
between the magnetic layers(Fig.~\ref{fig1}(a)). ST in asymmetric
nanopillars has been characterized by a distinct asymmetric
dependence of $dV/dI$ on $I$~\cite{cornellorig}. In contrast, our
samples show an approximately symmetric dependence of $dV/dI$ on $I$
(Fig.~\ref{fig1}(b)).  At $H=0$, increasing the magnitude of $I$
results in a gradual decrease of $dV/dI$ starting from $|I|\approx
I_c=1.6$~mA, as marked with dashed vertical lines. $dV/dI$ reaches a
minimum of $2.87$~$\Omega$ at $I=\pm4$~mA, followed by a small
increase at larger $I$. The minimum of $dV/dI$ is larger than the
P-state resistance $R_P=2.83$~$\Omega$, indicating that a stable P
state is not reached at any $I$. For $H>0$, large variations of
$dV/dI$ appear at $|I|>I_c$, stabilizing at $H>600$~Oe into gradual
increases. The data for $H\ge 800$~Oe only weakly depend on $H$. We
note a well-defined transition from the AP state at $|I|<I_c$ and
$H\le100$~Oe, to a P state at $H>100$~Oe, consistent with the MR
data of Fig.~\ref{fig1}(a). We attribute the more systematic
behaviors of our data compared to the results for Co-based
nanopillars~\cite{tsoisymmetric,kentsymmetric} to the smaller
dimensions of our nanopillars, indicated by a larger MR, and weaker
random crystalline anisotropy of Py. These factors decrease the
possibilities of inhomogeneous and noncollinear magnetic states.

Two distinct types of magnetic dynamics induced by ST have been
established for asymmetric nanopillars~\cite{myapl,cornellnature}.
At $H$ that was not too large, thermally activated flipping of the
magnetic moment resulted in sharp peaks and/or cusps in differential
resistance. ST-induced magnetic precession is induced at large $H$.
It is usually characterized by smooth, nearly independent of $H$
increases of MR. One may be tempted to attribute the large-$H$
behaviors of our symmetric samples to such precession. However, our
microwave spectroscopic measurements, as well simulations using the
Landau-Lifshits-Gilbert equations, show that coherent precession is
destroyed due to the ST-induced coupling of the dynamics of two
layers~\cite{myunpublished}. Instead, we focus here on the sharp
features observed in our samples at $|I|>I_c$ for $H<600$~Oe
(Fig.~\ref{fig1}(b)). Time-resolved measurements of resistance
showed that these features are caused by thermally activated
flipping of magnetic layers. The characteristics of the resulting
telegraph noise (TN) discussed below dramatically differ from the
usual properties of asymmetric samples~\cite{myprl}, enabling us to
unambiguously attribute them to the simultaneous effects of ST on
both magnetic layers.

\begin{figure}
\includegraphics[scale=1.05]{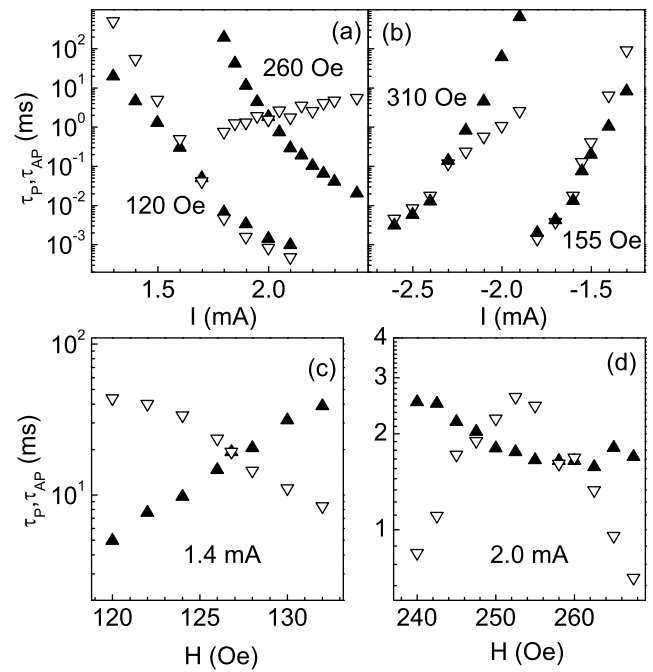}
\caption{\label{fig2} $\tau_P$ (solid symbols) and $\tau_{AP}$ (open
symbols) {\it vs} $I$ at fixed $H$ as labeled (a,b), and {\it vs}
$H$ at fixed $I$ as labeled (c,d). The uncertainties are similar to
the symbol sizes.}
\end{figure}

In the TN regime for asymmetric nanopillars at $I>0$, the average
dwell time $\tau_{P}$ in the P state  decreased with increasing $I$,
while the dwell time $\tau_{AP}$ in the AP state
increased~\cite{myprl}. Increasing $H$ had the opposite effect on
$\tau_{P}$ and $\tau_{AP}$. The results for symmetric nanopillars
are shown in Fig.~\ref{fig2}. We note that the MR in the TN
measurements are consistent with the static data of
Fig.~\ref{fig1}(a), indicating single domain behaviors. At a fixed
$H=120$~Oe, both $\tau_{P}$ and $\tau_{AP}$ decrease with increasing
$I>0$, remaining nearly equal over five orders of magnitude of their
variation (Fig.~\ref{fig2}(a)). These behaviors are not observed in
asymmetric nanopillars, and therefore must be attributed to the
effects of ST on both magnetic layers. Fig.~\ref{fig2}(a) also shows
that a different TN regime appears at a larger $H=260$~Oe. In this
regime, $\tau_{AP}$ increases with increasing $I$, while $\tau_{P}$
decreases. These latter behaviors can be easily explained when the
slight asymmetry of the nanopillar is taken into account. Large
enough $H$ must suppress the effects of ST on the larger F$_1$,
resulting in behaviors similar to those seen before in nanopillars
with a thick F$_1$.

Because of the symmetry of the sample, one can expect to see TN at
$I<0$ similar to that in Fig.~\ref{fig2}(a) for $I>0$. Indeed,
Fig.~\ref{fig2}(b) shows that the regime $\tau_{P}=\tau_{AP}$ is
also observed at $I<0$, although at a somewhat different $H=155$~Oe.
In the second regime at a larger $H=310$~Oe, both $\tau_{P}$ and
$\tau_{AP}$ still decrease, but at different rates following the
trend in Fig.~\ref{fig2}(a). Dependencies on $H$ shown in
Figs.~\ref{fig2}(c,d) explain the existence of two different TN
regimes. At small $H$ and $I>0$ (Fig.~\ref{fig2}(c)), $\tau_{P}$
increases, and $\tau_{AP}$ decreases with increasing $H$. These
trends become reversed twice at larger $H$ (Fig.~\ref{fig2}(d)). As
a result, the condition $\tau_{P}\approx\tau_{AP}$ is satisfied at
least twice when $H$ is increased. The complex nonmonotonic
variations of dwell times in Fig.~\ref{fig2} explain the large
irregular changes of MR in Fig.~\ref{fig1}(a): small (on the
logarithmic scale) relative changes of $\tau_{P}$ with respect to
$\tau_{AP}$ result in variations of $R$ between values very close to
$R_P$ and $R_{AP}$. These significant current-dependent variations
of $R$ appear as even sharper features in the differential
resistance $dV/dI$.

We interpret our results with a model for the dwell time of a
ferromagnet in the presence of ST~\cite{zhang}
\begin{equation}\label{ta}
\tau=\tau_0exp\left[\frac{E(H)(1-I/I_C)}{k_BT}\right],
\end{equation}
where $\tau$ is the dwell time in P or AP state, $\tau_0$ is attempt
time, $E(H)$ is the activation barrier at $I=0$, $I_C$ is the
critical current for the onset of current-induced dynamics at $T=0$,
and $k_B$ is the Boltzmann constant. For the following discussion,
we combine $k_BT/(1-I/I_C)$ into an effective current-dependent
temperature $T^*(I)$, which has a physically transparent
interpretation as describing the enhanced/suppressed by ST thermal
fluctuations of the magnetic moment.  Alternatively, one can
describe the effect of ST in terms of a current-dependent effective
barrier $E(H,I)$~\cite{krivorotov}. These two formally different
approaches equally well describe the nearly uniform dynamics of the
nanoscale magnetic layers.

\begin{figure}
\includegraphics[scale=0.6]{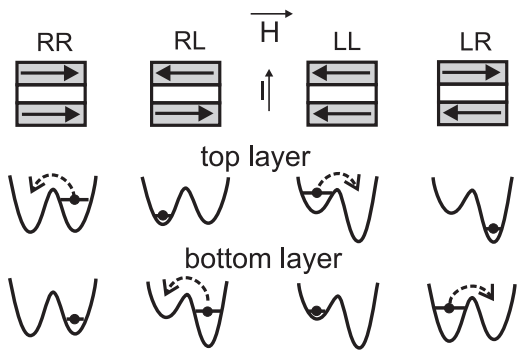}
\caption{\label{fig3} Four magnetic configurations of symmetric
nanopillars.  Lower panels are schematics for the activation
barriers and effective temperatures of the magnetic layers at $I>0$.
Dashed arrows show the transitions driven by ST, assuming that
effects of ST on the fluctuations of the two layers are
independent.}
\end{figure}

Assume that the effects of ST on the fluctuations of F$_1$ and F$_2$
are independent.  In this case, Eq.~\ref{ta} can be used to
separately determine the dwell times of each magnetic layer for
every possible configuration of the nanopillar. In the P state, the
critical current $I_{C1}$ describing the ST-driven excitation of
F$_1$ is positive, while the critical current $I_{C2}$ describing
the ST-driven excitation of F$_2$ is negative. In the AP state,
$I_{C1}<0$ and $I_{C2}>0$. According to Eq.~\ref{ta}, the effect of
ST at $I>0$ is to increase the effective temperature ${T^*}_{AP1}$
of F$_1$ in the AP state, thus decreasing the corresponding dwell
time $\tau_{AP1}$. Simultaneously, the effective temperature
${T^*}_{AP2}$ of F$_2$ is decreased, increasing its dwell time
$\tau_{AP2}$. As a result, F$_1$ flips with higher probability than
F$_2$, bringing the system into the P state. The effective
temperature ${T^*}_{P2}$ of F$_2$ in the P state is enhanced,
resulting in its subsequent flipping into AP state. The cycle of
sequential flipping of F$_1$ and F$_2$ is then repeated. By
symmetry, a similar cycle in reverse direction is expected for
$I<0$. This simplified model explains the simultaneous decreases of
$\tau_P$ and $\tau_{AP}$ with $I$ (Figs.~\ref{fig2}(a,b)). However,
analysis given below shows that it cannot fully account for the
dependence of data on $H$, indicating that the assumption of
independent effects of ST on F$_1$ and F$_2$ used in this model is
not fully justified.

Taking into account external and dipolar fields requires one to
consider four magnetic configurations denoted RR, RL, LL, and LR
based on the directions (Left or Right) of F$_1$ and F$_2$, as
illustrated in Fig.~\ref{fig3} for $I>0$ and $H\approx H_d$ pointing
to the right. Describing the dwell times of both layers in each
configuration by Eq.~\ref{ta} yields a sequence of transitions
RR-RL-LL-LR-RR, as shown by dashed arrows in the schematics of
activation energies. Based on this analysis, one expects
$\tau_{LL}\ll\tau_{RR}\approx\tau_{LR}\ll\tau_{RL}$ for the dwell
times in the corresponding configurations. This implies
$\tau_{AP}>\tau_P$, which is inconsistent with the regime
$\tau_P=\tau_{AP}$ in Fig.~\ref{fig2}. Moreover, increasing $H$
should increase both $\tau_{AP}$ which is dominated by the slow
transition RL-LL, and $\tau_{P}$ which is dominated by the slow
transition RR-RL. This explains the $H<254$~Oe part of panel (d),
but is inconsistent with the data of Fig.~\ref{fig2}(c). These
shortcomings of the model likely originate from the oversimplified
assumption that the effects of ST on the fluctuations of two layers
are independent. Indeed, considering the state RL, enhanced
fluctuations of F$_1$ due to ST must also induce fluctuations of
F$_2$ because ST exerted on both layers is similar in magnitude,
coupling their dynamics~\cite{slonczewski}. Since the activation
barrier for F$_1$ is smaller than for F$_2$, and decreases with $H$,
reversal of F$_1$ due to such coupling explains the decrease of
$\tau_{AP}$ in the data of Fig.~\ref{fig2}(c).

To summarize, we have demonstrated that simultaneous effects of ST
on both magnetic layers in symmetric nanopillars determine their
thermally activated behaviors. Our data indicate dynamical coupling
between the magnetic layers mediated by ST, which may have a
significant effect on the stability of magnetoelectronic devices and
the efficiency of ST for magnetic reversal or excitation of
dynamical states.

This work was supported by the NSF Grant DMR-0747609. NA and AH
acknowledge support from the NASA Space Grant Consortium.

\end{document}